\documentstyle[aps,prl,epsfig,amsmath,amssymb,multicol]{revtex}

\newcommand{\be}{\begin{eqnarray}}
\newcommand{\ee}{\end{eqnarray}}

\begin{document}

\title{Vortex Dynamics in a Coarsening Two Dimensional XY Model}
\author{Hai Qian and Gene F. Mazenko}
\address{James Franck Institute and Department of Physics, University of Chicago, Chicago, Illinois 60637}
\date{04/14/2003}
\maketitle

\begin{abstract}
The vortex velocity distribution function for a 2-dimensional 
coarsening 
non-conserved  $O(2)$ time-dependent Ginzburg-Landau model
is determined numerically and compared to theoretical predictions.
In agreement with these predictions the distribution
function scales with the average vortex speed which is
inversely proportional to $t^x$, where $t$ is
the time after the quench and $x$ is near
to $1/2$.
We find the entire curve, including a large speed algebraic tail, in good 
agreement with the theory.
\end{abstract}


\begin{multicols}{2}

\section{Introduction}

It is important to understand the role of defects in phase ordering
\cite{bray94} problems. 
We investigate here the growth kinetics of the non-conserved 
$O(2)$
symmetric time-dependent Ginzburg-Landau (TDGL) model in
two dimensions after a quench from a disordered high temperature state 
to zero temperature. The dominant structures in the ordering
kinetics of this system are vortices with charges $\pm 1$.
Vortices with higher order charges are unstable. Various aspects of the
defect structure have been explored in some detail before \cite{liu921,m90}.
We focus here on a numerical determination of the velocity distribution of the vortices
as a function of time $t$ after the quench.
Theory predicts \cite{maz97} that the distribution
function scales with the average vortex speed which is
inversely proportional to a length scale $L(t)$, which grows with time
$t$ after the quenches.
One also finds a large speed algebraic tail in good
agreement with predictions of an exponent of $-3$. In terms of the
velocity distribution this corresponds to an exponent of $-4$.
The number of vortices is also 
counted and its evolution in time is found to be consistent 
with previous work \cite{Y1993}.

The disordering agents in the phase ordering of the  
$n=d=2$ non-conserved TDGL
model ($2d$ XY model) are well known, where $d$ is the spatial dimension and
$n$ is the number of the components of the order
parameter.  One 
has
unit one charged vortices and, at nonzero temperature, spin waves.
If we focus on quenches to zero temperature we have only the vortices to 
consider.  Thus a typical vortex configuration is shown in Fig. 1.  As the time
evolves one has vortex anti-vortex annihilation until finally there are
no surviving vortices and the system is fully ordered (we only consider
the zero temperature case with no thermal fluctuations, where the system
does eventually order).  
For general $n=d$ Bray and Rutenberg \cite{br95} have shown that
the growth law for such systems is given by $L(t)\approx t^{1/2}$.
The exception is for $n=d=2$ where 
their method is mute. The growth law for this case was treated by Pargellis \textit{et
al.} \cite{P1992}, and checked numerically by Yurke \textit{et al.} \cite{Y1993}. 
There is a logarithmic correction  to  the scaling
law: $L(t)\approx (t/\log (t))^{1/2}$. 

\begin{figure} 
\begin{center}\includegraphics[scale=0.7]{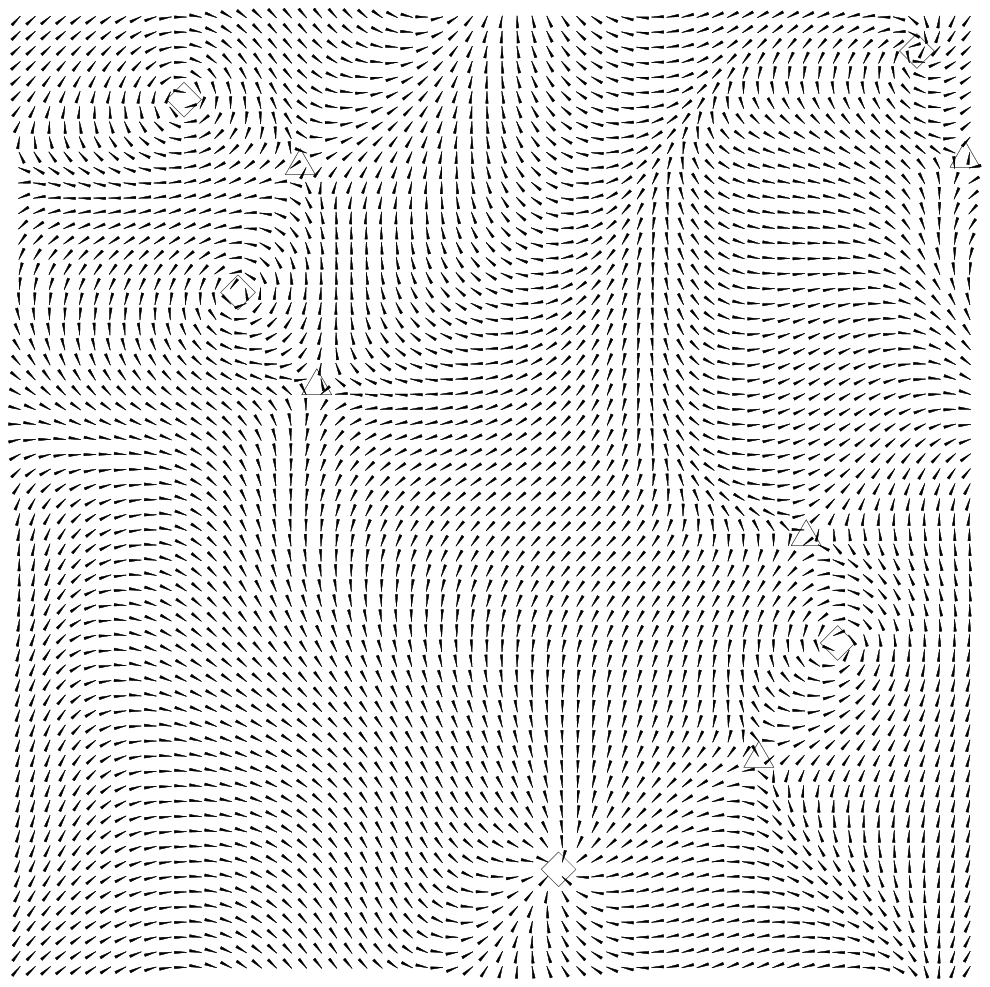}\end{center}
\caption{A typical vortex configuration in a $256\times 256$ 
system with lattice spacing $\Delta r=\pi/4$. The arrow on each each
site represents the
order parameter at that point. Not all the lattice sites are shown. The
squares and triangles are in the core regions of $+1$ and $-1$ vortices
respectively, where the magnitude of the order parameter is near
zero. The vortex core regions are picked out by using the method
described in the text.}
\end{figure}

There has also been theoretical work on the dynamics of these vortices.
Mazenko\cite{maz97} showed that if the order parameter can be assumed to
be a gaussian field \cite{liu92} when constrained to be near a vortex core then the vortex
velocity probability distribution, for $n=d$ \cite{maz97,maz99},
has the simple form:
\be
P({\vec v})\,d^nv=\frac{\Gamma (1+n/2)}{(\pi s^{2})^{n/2}}
\frac{1}{\left( 1+v^{2}/s^2\right)^{(n+2)/2}}\,d^nv\ ,
\label{eq:1}
\ee
where the scaling speed $s$ varies as $L^{-1}$
for long times. If we only care about the magnitude of the velocity and
integrate out the directions, then in the case of $n=d=2$ we have the
speed probability distribution
\begin{equation}
P(v)\,dv=\frac{2}{s^2}\,\frac{v}{\left(1+v^2/s^2\right)^2}\,dv\ ,
\nonumber
\end{equation}
or equivalently
\begin{equation}
P({\widetilde v})\,d{\widetilde v}=\frac{2a\,{\widetilde v}}{(1+a{\widetilde v}^2)^2}\,d{\widetilde v}\ ,
\end{equation}
with $a =(\pi/2)^2$ and ${\widetilde v}=v/\bar{v}$, where the average speed $\bar{v} = \pi s/2$. So, after being scaled with the average speed, the vortex
speeds have the same probability distribution at different times.
A key feature of the predictions for $P(v)$ is that there is  a 
large velocity algebraic tail $\sim v^{-3}$.  This
tail was also found using scaling arguments by Bray\cite{bray97}.
In this paper we check these predictions numerically for the case $n=d=2$.
We find that velocity
probability distribution function does obey scaling of the form predicted by
Eq. (2). But the average speed falls off as $t^{-1/2}$ without the
logarithmic correction and there is a
large speed tail consistent with an exponent of $-3$.

We have also monitored the vortex density $n_{v}$ as a function of time
and find, in agreement with previous work\cite{Y1993}, 
\be
n_{v}\propto L_{v}^{-2}(t)
\label{eq:2}
\ee
where $L_{v}(t)\propto (t/\log (t))^{1/2}$. The vortex number density
should be proportional to the system's  energy above its ground energy
$E_0$ ($=-\epsilon^2/4$) per unit area, where $\epsilon$ is the control
parameter of the system (see Eq. (4) below). Our
simulation verifies this result.

\section{System Description}

The system we study is described by the Langevin equation
\begin{equation}
\frac{\partial \psi_i}{\partial t}=\epsilon \psi_i+\nabla^2
\psi_i-\left(\vec{\psi}\right)^2 \psi_i\ ,
\end{equation}
where $i=1,2$ are the indices for the two components of the order
parameter $\vec{\psi}$. The noise term is zero because the system is
quenched to zero temperature. By choosing proper units for the time and
space and rescaling the order parameter,  $\epsilon$ can take on any
positive value. The equation is put on a square lattice with periodic
boundary conditions and driven by
the finite difference scheme, i.e. replacing 
$\partial_t \psi_i({\bf r},t)$ by
$\left[\psi_i^{m+1}(kl)-\psi_i^{m}(kl)\right]/\Delta t$ with $m$ being
the time step number, and $\nabla^2
\psi_i({\bf r},t)$ by 
\begin{equation}
\nabla^2\psi_i(kl)=\frac{1}{(\Delta
r)^2}\left[\frac{2}{3}\sum_{NN}+\frac{1}{6}\sum_{NNN}-\frac{10}{3}\right]\psi_i(
kl)
\end{equation}
where $NN$ and $NNN$ mean the nearest neighbors and next-nearest
neighbors respectively and the lattice point is  ${\bf r}/\Delta r=(k,l)$. Here $\Delta t$ is the time step and $\Delta r$
is the lattice spacing. Both are dimensionless.

We have studied two systems in some detail. In both we choose $\epsilon =0.1$ and use $1024\times 1024$  lattice sites.  
In system one, or the bigger system, we use $\Delta t=0.02$ and 
$\Delta r=\pi/4$.  In system two, or the smaller system, we use $\Delta t=0.01$
and $\Delta r=\pi/8$. In both we measured the vortices number and the
system energy. We measured the vortex
speed distribution only in the bigger system.

We prepare the system initially in a completely disordered state. The
average magnitude $\bar{M}$ of the vector order parameter $\vec{\psi}$ at time $t$ is
calculated and the vortices' core regions are identified with those sites
on which the order parameter magnitudes
$|\vec{\psi}|<\bar{M}/4$. Usually each core has about 10 sites in it. Here the coefficient $1/4$ is appropriately chosen
so that no vortices are missed and no irrelevant points are picked. 
A circular integration around each
vortex produces $2\pi$ or $-2\pi$, which corresponds to two types of
vortices of topological charges $+1$ and $-1$ respectively. In fact
there are situations where we obtain charge $0$. This is due to the non-zero area of
the integration circle. When a pair of $+1$ and $-1$ vortices
annihilate and the distance between them becomes smaller than the size
of the integration circle, the circular integration will reflect the sum of
the two
charges, which is $0$. However, the lifetime
of these $0$ charges are much smaller than the lifetime of the $\pm 1$
vortices. So they do not affect our statistics.
We find in our simulations that the numbers of positive and negative
vortices are equal.

The position of a vortex is given by the center of its core
region. Suppose the order parameter's magnitude at the core region is
described by $M(x_i, y_i)$ with $(x_i,y_i)$ belonging to the core
region. Then by fitting $M(x_i,y_i)$ to the function
$M(x,y)=A+B[(x-x_0)^2+(y-y_0)^2]$ we can find the center $(x_0,y_0)$.

The positions of 
each vortex at different times are recorded, and the speed is calculated 
using $v=\Delta d/\Delta \tau$. Here $\Delta d$ is the distance that the 
vortex travels in time $\Delta \tau$.  We have, simultaneously,
recorded the number of vortices as a function of time and checked the
scaling result given by Eq. (\ref{eq:2}).

\section{Numerical Results}

Every $\delta t=10$ we compute the speed of each vortex with $\Delta
\tau=5$, i.e. $v=|{\bf r}(t+\Delta \tau)-{\bf r}(t)|/\Delta \tau$ with
$t$ increasing by step length $\delta t$ and ${\bf r}$ being the
position of the vortex.
We found the average speed of the vortices $\bar{v}(t)$ is
proportional to $t^{-x}$ with $x=0.51\pm 0.01$ for the bigger system ($\Delta
r=\pi/4$)  as
shown in Fig. 2. Unlike the case for $L_v(t)$
extracted from $n_v$ and $(E-E_0)$, we do not observe logarithmic
correction for $\bar{v}(t)$. We can not rule out such corrections
appearing on a longer time scale but they do not enter on the same time
scale as for $L_v(t)$.

\begin{figure} 
\begin{center}\includegraphics[scale=0.33]{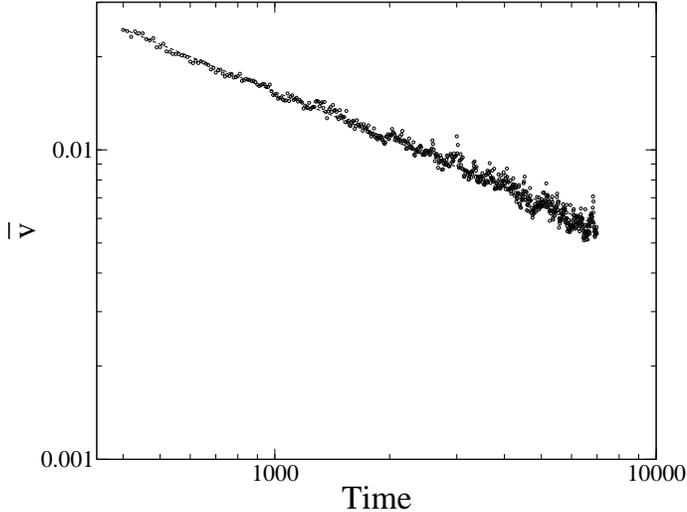}\end{center}
\caption{The average speed $\bar{v}(t)$ of the vortices for
the bigger system ($\Delta r=\pi/4$). The speed
of each vortex is calculated with $\Delta \tau=5$. $t^{-x}$ is used to fit 
the data. $x=0.51\pm 0.01$ for the system with $\Delta r=\pi/4$. The data are 
averaged over $60$ different initial conditions.}
\end{figure}

We compared the speed distribution at different
times. They have approximately the same shape after being rescaled by the
average speed. At early times this is clear. At late times there are
fewer data points and the
similarity between the two distributions at different times is not so
apparent. Even at early times the number of data points are not enough
to give a good fit to the distribution's long tail. So 
we rescale
the speed data with the best fit to the average speed $\bar{v}(t)\propto
t^{-0.51}$ and put the data 
for all times into one
histogram as shown in Fig. 3. By this means we obtain better statistics. 
We find that the distribution can be well fit to the function
\begin{equation}
P({\widetilde v}) = \frac{2a{\widetilde v}}{(1+a {\widetilde v}^2)^2}\ ,
\end{equation}
for $a$ between 2 and 2.5. The best fit is for $a = 2.12$. However if we
require that the average speed is given by the measured speed then we
must have $a=(\pi/2)^2= 2.47$.  The best fit
and the most consistent fit are both shown in Fig. 3. 
The distribution has a long tail which 
is approximately $(v/\bar{v})^{-y}$ with $y=3$. If we just try to fit
the long tail, then the best fit is given by $a=2.47$. These results are in
excellent quantitative agreement with the theoretical prediction in
Ref. \cite{maz97}. 

\begin{figure} 
\begin{center}\includegraphics[scale=0.32]{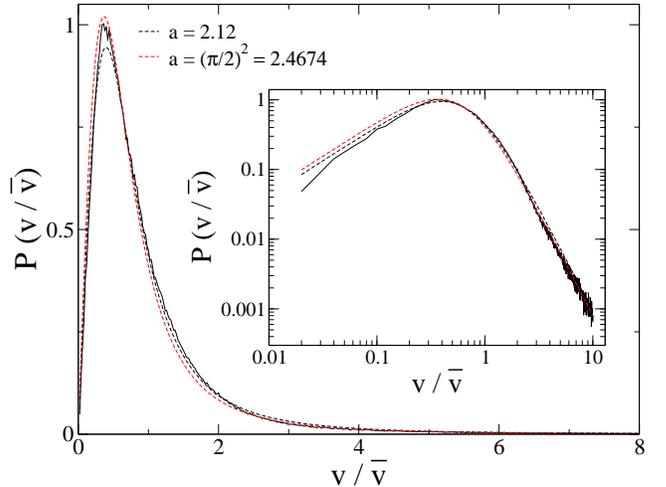}\end{center}
\caption{The vortices' speed distribution probability density for the
$\Delta r=\pi/4$ system $P(v/\bar{v}(t))$ (the solid line) fit to the function
$P(x)=2ax/(1+ax^2)^2$ with 
$a=2.12$ and $x=v/\bar{v}(t)$. The theoretical curve with $a=(\pi/2)^2$
is also shown in the figure. The large speed tail of the distribution 
can be fit to $t^{-y}$ with $y=3$. The insert shows the same data
in the logarithmic scale, and it seems that the the theoretical curve fits
the tail better. The data  are 
averaged over $60$ different
initial conditions. In the calculation of the probability
distribution we use the bin width $0.02$.}
\end{figure}

In these results we have scaled the velocity with $\sim t^{-z}$ by taking $z=x$ just as
the theory predicts. However
the value of scaling exponent $z$ of the speed distribution is not very
robust in our simulations. If we change the exponent $z$ and use $t^{-z}$ to rescale the speed
distribution at different times, and they also have approximately the
same shape.  So the uncertainty of $z$ is quite large. After being
rescaled by $t^{-z}$ and put into one histogram, the speed distribution
for small and intermediate values of ${\widetilde v}$ are somewhat
dependent on the value of $z$.  However, the power-law
tail is quite robust. We alter the value
of $z$ between $0.4$ and $0.6$, and find that the tail exponent changes
by less than 0.1. 

\begin{figure} 
\begin{center}\includegraphics[scale=0.31]{figs/fig4.eps}\end{center}
\caption{The total number of the vortices per unit area $n_v(t)/S$ times $t$, where
$S=(1024\times\Delta r)^2$. In both systems $n_v\cdot t/S$ can be fit
to $a \log(t/t_0)$. The data from the large and the small systems are
averaged over $68$ and $58$ different initial conditions
respectively.}
\end{figure}

\begin{figure} 
\begin{center}\includegraphics[scale=0.31]{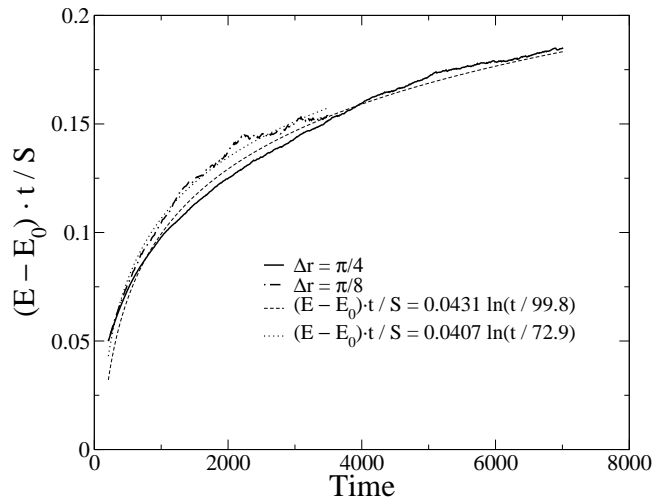}\end{center}
\caption{The product of the systems energy per unit area $(E-E_0)/S$ with
$t$. The ground state
energy is $E_0=-\epsilon^2S/4$. The data can also be fit to $a
\ln(t/t_0)$. The data from the large and the small systems are
averaged over $68$ and $58$ different initial conditions
respectively.}
\end{figure}

In both the larger and smaller systems the vortex number densities have the same time dependence. We obtain
$n_{v}\sim \left[t/\log(t/t_0)\right]^{-1}$, where $n_{v}$ is the number
density of
vortices (positive or negative) and $t_0$ is a constant. In Fig. 4, we
show the data for $n_{v}\cdot t/S$. $S$ is the area of the system.
In Fig. 5, we show the data for the energy per
unit area $(E-E_0)/S$ times $t$. We conclude that the energy density is proportional to
the number density of
the vortices.

\section{Conclusions}

We have studied the growth kinetics of the non-conserved TDGL model in
the case of $n=d=2$. We measured the speed probability distribution for
the $\pm 1$ vortices. At any given time with relatively few vortices, the statistics are poor. However the accumulated data for all
times when scaled gives a scaling function with good
statistics. Although the scaling exponent $z$ is 0.5 with significant
uncertainty, the large speed tail does takes the form of
$(v/\bar{v})^{-y}$ with the exponent $y=3$. This is consistent
with the theoretical prediction. The form of the distribution
$P({\widetilde v})$ is quantitatively consistent with the theoretical
prediction. 

Why does the theory do so well? It was shown by Mazenko and Wickham
\cite{maz98}, that one can construct a nontrivial self-consistent
gaussian theory
for the order parameter if it is constrained to be evaluated near
a vortex core. Such constraints occur naturally, for example, in
averages over the vortex density. This suggests that the theory developed in Ref. \cite{maz97} may be on
a firmer footing than first thought.

According to the theory, the average speed should be proportional to the
inverse of the correlation length $L(t)$. We did not observe any
logarithmic correction in the average speed, although it appears in the
correlation length when $n=d=2$. However a logarithmic behavior may
still exist. The time we used in our simulations may not be long enough
to see the effect.

We also measured the number density of the vortices and the energy
density of the two systems. Their time dependences both have a
logarithmic correction, which is consistent with previous work. 

\vspace{5mm}

Acknowledgments: This work was supported by the National Science Foundation
under Contract No. DMR-0099324.

\end{multicols}

\end{document}